\documentclass[twocolumn]{jpsj2} %% two-column layout
\title{Spin Frustration and Orbital Order in Vanadium Spinels}

\author{Yukitoshi \textsc{Motome}$^{1}$\thanks{E-mail: motome@riken.jp} and Hirokazu \textsc{Tsunetsugu}$^{2}$}

\inst{$^{1}$RIKEN (The Institute of Physical and Chemical Research), Saitama 351-0198, Japan\\
$^{2}$ Yukawa Institute for Theoretical Physics, Kyoto University, Kyoto 606-8502, Japan}

\abst{
We present the results of our theoretical study on the effects of geometrical frustration 
and the interplay between spin and orbital degrees of freedom 
in vanadium spinel oxides $A$V$_2$O$_4$ ($A$ = Zn, Mg or Cd). 
Introducing an effective spin-orbital-lattice coupled model 
in the strong correlation limit and 
performing Monte Carlo simulation for the model, 
we propose a reduced spin Hamiltonian in the orbital ordered phase 
to capture the stabilization mechanism 
of the antiferromagnetic order. 
Orbital order drastically reduces spin frustration 
by introducing spatial anisotropy in the spin exchange interactions, and 
the reduced spin model can be regarded as 
weakly-coupled one-dimensional antiferromagnetic chains.
The critical exponent estimated by finite-size scaling analysis 
shows that the magnetic transition belongs to 
the three-dimensional Heisenberg universality class. 
Frustration remaining in the mean-field level is reduced  
by thermal fluctuations to stabilize a collinear ordering. 
}

\kword{vanadium spinel oxides, pyrochlore lattice, 
geometrical frustration, $t_{2g}$ electrons, 
Kugel-Khomskii model, orbital ordering, antiferromagnetic ordering, 
Heisenberg universality class, thermal fluctuation, order-by-disorder mechanism}

\begin{document}
\maketitle

\section{Introduction} 

Spinels are one of the most typical geometrically-frustrated systems. 
Among them, the so-called $B$ spinel oxides $AB_2$O$_4$, 
where $A$ cations are nonmagnetic, have attracted much interests 
since magnetic $B$ cations form the pyrochlore lattice, which consists of 
a three-dimensional (3D) network of corner-sharing tetrahedra. 
The system suffers from strong magnetic frustration: 
Magnetic correlations are strongly suppressed 
and complex phenomena may appear at low temperatures 
due to a large number of nearly-degenerate ground states. 

In this paper, we will investigate typical $B$ spinels
$A$V$_2$O$_4$ with divalent nonmagnetic $A$ cations such as Zn, Mg or Cd. 
Besides the geometrical frustration,
these vanadium spinels have another key issue, 
the orbital degree of freedom. 
Since each V$^{3+}$ cation has two $3d$ electrons in threefold $t_{2g}$ levels, 
the system has the orbital degree of freedom in addition to the spin. 
Hence, vanadium spinel oxides are intriguing systems 
which provide two important issues in strongly correlated systems, 
i.e., geometrical frustration and the interplay 
between spin and orbital degrees of freedom. 

\begin{figure}[t]
\begin{center}
\includegraphics[width=8cm]{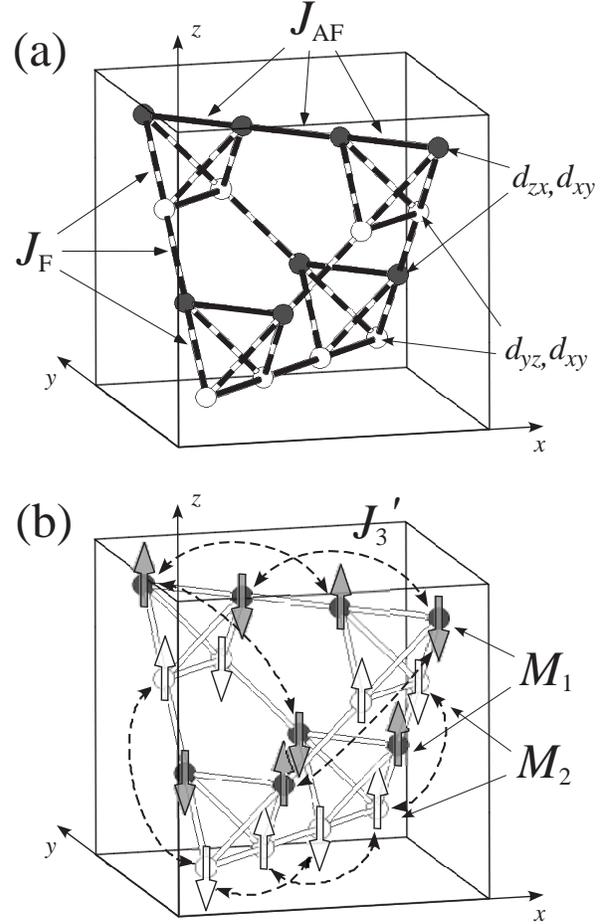}
\end{center}
\caption{(a) Effective spin exchange interactions for nearest neighbors
in the orbital ordered phase in a cubic unit cell of the pyrochlore lattice. 
White (gray) circles are the V sites
where the $d_{yz}$ and $d_{xy}$ ($d_{zx}$ and $d_{xy}$) orbitals are occupied.
Solid (dashed) bonds represent the strong AFM interactions $J_{\rm AF}$ 
(the weak FM interactions $J_{\rm F}$) 
in the $xy$ ($yz$ and $zx$) planes.
(b) Effective spin exchange interactions for third neighbors, $J_3'$. 
Two sublattices are shown which are connected by $J_3'$.}
\label{fig1}
\end{figure}

Vanadium spinels $A$V$_2$O$_4$ ($A$ = Zn, Mg or Cd) 
show two different phase transitions 
at low temperatures; 
one occurs at around $50$K and the other is at around $40$K. 
\cite{Ueda1997}
Note that the transition temperatures are significantly lower than 
Curie-Weiss temperature $\sim 1000$K, 
\cite{Muhtar1988}
which is attributed to the effect of geometrical frustration. 
The former is a structural transition from 
high-temperature cubic phase to low-temperature tetragonal phase. 
The latter is a magnetic transition 
with a collinear antiferromagnetic (AFM) ordering. 
The magnetic structure consists of the staggered ordering with 
$\uparrow$-$\downarrow$-$\uparrow$-$\downarrow$-$\cdots$ structure 
along the $xy$ chains and
the period-four ordering with 
$\uparrow$-$\uparrow$-$\downarrow$-$\downarrow$-$\cdots$ structure 
along the $yz$ and $zx$ chains 
\cite{Niziol1973}
[see Fig.~\ref{fig1} (b)]. 
(We take the $z$ axis in the tetragonal $c$ direction.)
The issue is the microscopic mechanism of the two transitions: 
How is the degeneracy due to the geometrical frustration lifted? 
What is the role of the orbital degree of freedom?

The authors have proposed an effective spin-orbital-lattice model 
for this problem and shown that the two transitions are well reproduced 
by Monte Carlo (MC) simulation performed for this model.
\cite{Tsunetsugu2003,MotomePREPRINT}
In the present study, we will focus on the magnetic frustration
in the spin-orbital coupled system. 
We will introduce a reduced spin model to explain 
the low energy physics in the orbital ordered phase, and 
clarify the nature of the magnetic transition 
including its critical properties. 

This paper is organized as follows. 
In Sec.~2, we introduce the effective spin-orbital-lattice model 
which has been derived in the previous publications, 
\cite{Tsunetsugu2003,MotomePREPRINT}
and describe the method of calculations. 
In Sec.~3, we propose a reduced spin Hamiltonian in the orbital ordered phase 
and explain the microscopic mechanism of stabilizing the complex AFM ordering. 
Numerical analysis of the critical exponent is also shown. 
Section 4 is devoted to summary.

\section{Model and Method} 

We start from 
a multiorbital Hubbard model with three $t_{2g}$ orbitals 
on the pyrochlore lattice, and 
consider the perturbation in the strong correlation limit to describe 
the low energy physics of the insulating vanadium spinels. 
By using atomic eigenstates 
with two $3d$ electrons per site 
in a high-spin state as the unperturbed states,  
we derive the effective spin-orbital-lattice coupled model 
(the so-called Kugel-Khomskii type Hamiltonian\cite{Kugel1973}) 
in the form 
\cite{Tsunetsugu2003,MotomePREPRINT}
\begin{eqnarray}
&& H = H_{\rm so} + H_{\rm JT}, 
\label{eq:H}
\\
&& H_{\rm so} = -J \sum_{\langle ij \rangle} h_{ij} 
- J_3 \sum_{\langle\!\langle ij \rangle\!\rangle} h_{ij}, 
\label{eq:H_SO}
\\
&& h_{ij} = (A + B \mib{S}_i \cdot \mib{S}_j) 
[ n_{i \alpha(ij)} \bar{n}_{j \alpha(ij)} 
+ \bar{n}_{i \alpha(ij)} n_{j \alpha(ij)} ]
\nonumber \\
&& \quad \quad + C (1 - \mib{S}_i \cdot \mib{S}_j)
n_{i \alpha(ij)} n_{j \alpha(ij)}, 
\\
&& H_{\rm JT} = \gamma \sum_i Q_{i} \epsilon_i
+ \sum_i Q_i^2 / 2 
- \lambda \sum_{\langle ij \rangle} Q_i Q_j, 
\end{eqnarray}
where $\mib{S}_i$ is the $S=1$ spin operator and  
$n_{i \alpha}$ is the density operator for site $i$ and 
orbital $\alpha = 1$ ($d_{yz}$), $2$ ($d_{zx}$), $3$ ($d_{xy}$). 
Here, $\bar{n}_{i\alpha} = 1 - n_{i\alpha}$ and 
we impose a local constraint $\sum_{\alpha=1}^3 n_{i \alpha} = 2$ at each site. 
The summations with $\langle ij \rangle$ and 
$\langle\!\langle ij \rangle\!\rangle$ 
are taken over the nearest-neighbor (NN) sites and third-neighbor sites, respectively. 
Here, we take into account only the dominant $\sigma$-bond hopping integrals 
in the original multiorbital Hubbard model, for instance, 
hoppings between $d_{xy}$ orbitals in the same $xy$ plane. 
\cite{Tsunetsugu2003,MotomePREPRINT}
This approximation results in 
the orbital diagonal interaction in $H_{\rm so}$; 
$\alpha(ij)$ is the orbital which gives rise to the $\sigma$ bond 
between sites $i$ and $j$. 
$H_{\rm JT}$ describes the orbital-lattice coupling part, 
where $\gamma$ is the electron-phonon coupling constant 
of the tetragonal Jahn-Teller (JT) mode, 
$Q_i$ denotes the amplitude of local lattice distortion at site $i$, and 
$\epsilon_i = n_{i1} + n_{i2} - 2 n_{i3}$. 
$\lambda$ describes the interaction between NN JT distortions, 
which mimics the cooperative aspect of the JT distortion. 

An important feature of the model (\ref{eq:H}) is 
the symmetry of the intersite interactions. 
In the spin part, the interaction is isotropic Heisenberg type. 
On the other hand, in the orbital part, 
it depends only on the density operator, that is, 
there is no transverse component which mixes different orbitals. 
Hence, the orbital interaction is three-state Potts type 
corresponding to three $t_{2g}$ states.
Moreover, it depends on bond direction and 
also orbital states at both ends of a bond. 
These peculiar properties of the orbital interactions
are crucial to trigger the orbital ordering by lifting the degeneracy 
in the present frustrated system.
\cite{Tsunetsugu2003,MotomePREPRINT}

The parameters in $H_{\rm so}$ are given by the coupling constants 
in the starting multiorbital Hubbard Hamiltonian, and 
in the following we use the reasonable estimates 
given in Ref.~\ref{Motome} as 
$J_3/J = 0.02$ with $J \simeq 200$K, 
$A = 1.21$, $B = 0.105$, and $C = 0.931$.
For the JT parameters, we take $\gamma^2/J = 0.04$ and $\lambda/J = 0.15$, 
which are typical values to have the tetragonal distortion 
consistent with the experimental result. 
\cite{MotomePREPRINT}
Hereafter, we will set the lattice constant of the cubic unit cell 
as a length unit and 
use the convention of the Boltzmann constant $k_{\rm B}=1$.

We have studied thermodynamic properties of the model (\ref{eq:H})
by employing classical Monte Carlo (MC) simulation
to avoid the negative sign problem 
due to the geometrical frustration of the pyrochlore lattice.
Since quantum nature exists only in the spin $S=1$ operators in $H_{\rm so}$, 
we approximate them by classical vectors with unit modulus. 
\cite{MotomePREPRINT}. 
We typically perform $10^5$ MC samplings for measurements
after $10^5$ steps for thermalization.
System sizes are up to $L^3 = 12^3$ in the cubic units, 
which includes $12^3 \times 16$ vanadium sites. 

\section{Results} 

\subsection{Reduced spin model under orbital ordering}

MC results show that as temperature decreases, 
first an orbital ordering takes place 
with the tetragonal JT distortion, 
corresponding to the structural transition in experiments.
\cite{MotomePREPRINT}
This is triggered by the highly anisotropic nature of 
the three-state Potts-type orbital interaction described above.
This transition is discontinuous, and the orbital moment shows a large jump 
at the transition temperature $T_{\rm O} \simeq 0.19J$. 
The ordering structure is $A$ type, i.e., the layered order
with alternative stacking of ($d_{zx}$, $d_{xy}$) and
($d_{yz}$, $d_{xy}$) occupied planes in the $z$ direction 
as shown in Fig.~\ref{fig1} (a).

Since the orbital polarization below $T_{\rm O}$ is large and 
quickly converges to the saturated value, 
\cite{MotomePREPRINT}
it is convenient to consider a reduced spin Hamiltonian 
by freezing the orbital degree of freedom 
in order to capture the low energy physics in the orbital ordered phase.
The reduced spin Hamiltonian is obtained by replacing 
the orbital parts in Eq.~(\ref{eq:H_SO})
by their mean values in the fully polarized state.
For instance, we replace $n_{i3} n_{j3}$ and $n_{i3} \bar{n}_{j3}$ in $H_{\rm so}$ 
by $\langle n_{i3} n_{j3} \rangle = 1$ and 
$\langle n_{i3} \bar{n}_{j3} \rangle = 0$, respectively, 
for the bonds in the $xy$ planes.
Then we end up with the reduced model in the form
\begin{equation}
H_{\rm spin} 
= \! 
\sum_{\langle ij \rangle \in xy} \!\! J_{\rm AF} \mib{S}_i \cdot \mib{S}_j
+ \sum_{\langle ij \rangle \in yz,zx} \!\!\!\! J_{\rm F} \mib{S}_i \cdot \mib{S}_j
+ 
{\sum_{\langle\!\langle ij \rangle\!\rangle}}' J_3' \mib{S}_i \cdot \mib{S}_j.
\label{eq:H_spin}
\end{equation}
The first two terms describe the NN spin exchange interactions, in which  
the former (latter) summation is taken over 
the NN pairs in the $xy$ ($yz$ and $zx$) planes. 
$J_{\rm AF} = JC > 0$ is antiferromagnetic (AFM) and 
$J_{\rm F} = -JB < 0$ is ferromagnetic (FM). 
These exchange interactions are shown in Fig.~\ref{fig1} (a). 
The last term describes the third-neighbor interactions, 
where the summation is taken over the third-neighbor $\sigma$ bonds 
shown in Fig.~\ref{fig1} (b). $J_3' = J_3 C >0$ is AFM.
Here we omit the orbital-lattice part for simplicity.

First we consider only the NN exchange interactions. 
One important point is the magnitude 
of two different interactions $J_{\rm AF}$ and $J_{\rm F}$. 
By using the parameters in Sec.~2, 
we obtain $J_{\rm AF} \simeq 0.931J$ and 
$|J_{\rm F}| \simeq 0.105J$, and
hence the AFM exchange in the $xy$ planes is about 
ten times stronger than the FM exchange 
in the $yz$ and $zx$ planes. 
Thus, the system can be regarded as 
one-dimensional (1D) AFM chains coupled by weak FM exchange interactions.
Another key feature is the frustration 
due to the interchain coupling $J_{\rm F}$. 
Once the AFM correlation develops along the $xy$ chains by the strong $J_{\rm AF}$, 
coupling of the $xy$ chains are frustrated; 
in the mean-field level, the total energy does not depend on 
the relative angle of AFM moments in any two $xy$ chains. 
Therefore, the system is reduced to independent 1D AFM chains and 
it is hard to establish a three-dimensional (3D) AFM order at this stage. 

The frustration remaining among the $xy$ chains is almost reduced 
by the small third-neighbor exchange $J_3' \simeq 0.0186 J$.
As shown in Fig.~\ref{fig1} (b), 
$J_3'$ connects the parallel $xy$ chains and 
stabilizes a 3D AFM ordering. 
Note that however there still remains 
frustration between two sublattices; 
one consists of ($d_{zx}$, $d_{xy}$) occupied sites and
the other consists of ($d_{yz}$, $d_{xy}$) occupied sites. 
The relative angle between two sublattice moments 
$\mib{M}_1$ and $\mib{M}_2$ is free in the mean-field level. 
Reduction of the frustration will be discussed in Sec.~3.3.

\begin{figure}[t]
\begin{center}
\includegraphics[width=7cm]{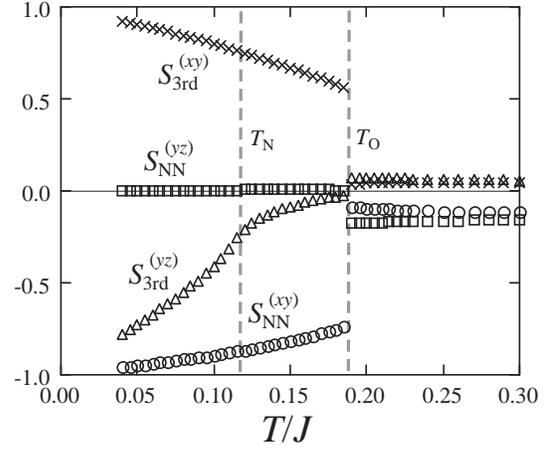}
\end{center}
\caption{
Temperature dependences of spin correlations 
for the system size $L=12$. 
Error bars are smaller than the symbol sizes. 
See the text for details.
}
\label{fig2}
\end{figure}

The picture based on the reduced spin model (\ref{eq:H_spin}) is 
confirmed by our MC calculations. 
Figure~\ref{fig2} shows the temperature dependences of 
the NN and third-neighbor spin correlations 
in the effective spin-orbital-lattice model (\ref{eq:H}): 
$S_{\rm NN (3rd)}^{(\nu)} = \sum_{\langle ij \rangle 
(\langle\!\langle ij \rangle\!\rangle) \in \nu} 
\langle \mib{S}_i \cdot \mib{S}_j \rangle / N_{\rm b}$, 
where $N_{\rm b}$ is the number of bonds in the summation and 
$\nu = xy, yz, zx$.  
Note that $S^{(yz)}$ equals to $S^{(zx)}$ by symmetry. 
Below the orbital ordering temperature $T_{\rm O}$, 
AFM correlations develop along the $xy$ chains 
($S_{\rm NN}^{(xy)} < 0$ and $S_{\rm 3rd}^{(xy)} > 0$), 
while correlations in the $yz$ and $zx$ chains remain small. 
Particularly, the NN correlations $S_{\rm NN}^{(yz)}$ become almost zero 
because of the frustration in the interchain coupling $J_{\rm F}$. 
Below $T_{\rm N} \simeq 0.12J$ (which will be assigned to 
the 3D AFM transition temperature in Sec.~3.2), 
$S_{\rm 3rd}^{(yz)}$ rapidly grows due to $J_3'$. 
Thus the system becomes highly one-dimensional
below $T_{\rm O}$, and 
3D interchain correlations develop below $T_{\rm N}$. 

Consequently, the orbital ordering introduces
the spatial anisotropy in the intersite spin exchange interactions, 
which plays a key role to reduce the geometrical frustration.
The system is effectively reduced to weakly-coupled 1D chains and 
finds a way to establish a 3D AFM order.

\subsection{Magnetic transition and universality class}

\begin{figure}[t]
\begin{center}
\includegraphics[width=8.5cm]{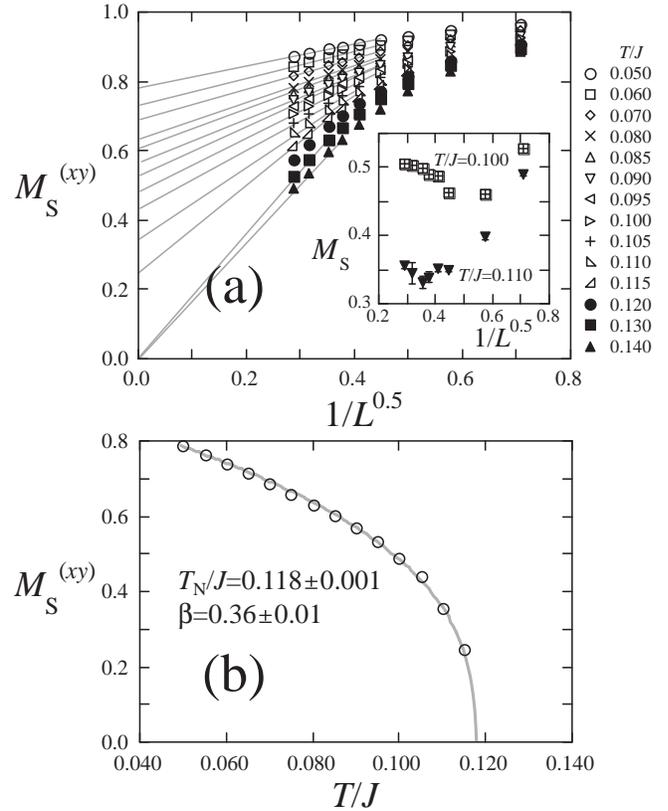}
\end{center}
\caption{(a) System size extrapolation of the staggered moment 
along the $xy$ chains. Lines are the linear fits for the data.
Inset: System size dependence of $M_{\rm S}$.
(b) Temperature dependence of the extrapolated data in (a).
The curve shows the fit by $M_{\rm S}^{(xy)} \propto (T_{\rm N} - T)^\beta$. 
Error bars are smaller than the symbol sizes.}
\label{fig3}
\end{figure}

The 3D AFM order is indeed found 
to occur in our MC results for the model (\ref{eq:H}).  
The staggered magnetic moment $M_{\rm S}$, 
which is the magnitude of the staggered moment $\mib{M}_1$ or $\mib{M}_2$, 
grows continuously at $T \sim 0.12J < T_{\rm O}$.
To determine the transition temperature $T_{\rm N}$ precisely, 
we perform a finite-size scaling analysis.
As shown in the inset of Fig.~\ref{fig3} (a), 
$M_{\rm S}$ shows non-monotonic size dependence 
within the range of the system sizes calculated here, 
probably because of the large spatial anisotropy 
of the AFM correlations discussed in Sec.~3.1.
Instead, we use the staggered moment along the AFM $xy$ chains, 
$M_{\rm S}^{(xy)}$, 
which is defined by the average of the magnitude of AFM moments 
in the $xy$ chains. 
\cite{MotomePREPRINT}
The restricted summation reduces 
effects of anisotropic correlations.

Fig.~\ref{fig3} (a) shows the system-size dependences of $M_{\rm S}^{(xy)}$.
The MC data are fitted by a linear function of $1/\sqrt{L}$ 
(Ref.~\ref{Motome}). 
The values extrapolated to $L \to \infty$
are plotted as a function of temperature in Fig.~\ref{fig3} (b).

The transition temperature $T_{\rm N}$ and the critical exponent $\beta$
are obtained by the scaling fit 
$M_{\rm S}^{(xy)} \propto (T_{\rm N} - T)^\beta$
for the data in Fig.~\ref{fig3} (b). 
The best fit gives the estimates 
$T_{\rm N}/J = 0.118 \pm 0.001$ and $\beta = 0.36 \pm 0.01$.
The exponent $\beta$ is consistent with 
that of the 3D isotropic Heisenberg model 
with short-range interactions, $\beta=0.365$ 
(Ref.~\ref{LeGuillou}), 
which indicates that the AFM transition in the present system 
belongs to the 3D Heisenberg universality class.

\subsection{Collinear ordering due to `order by disorder'}

\begin{figure}[t]
\begin{center}
\includegraphics[width=7.5cm]{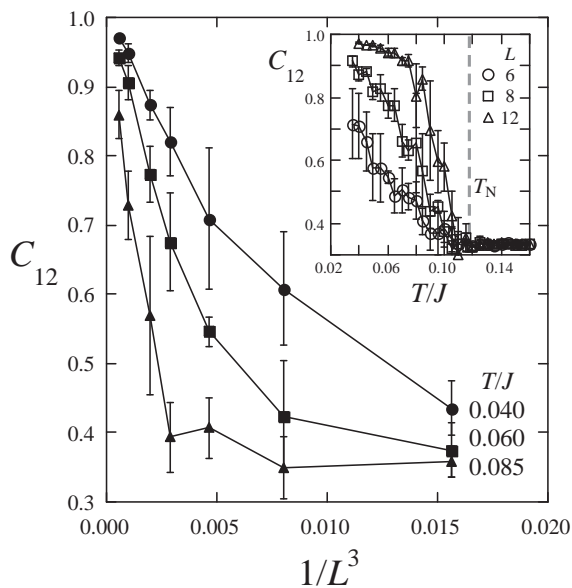}
\end{center}
\caption{(a) Temperature dependence of the collinearity.
(b) System size dependence. The lines are guides for the eyes.}
\label{fig4}
\end{figure}

As explained in Sec.~3.1, in the mean-field level,
frustration remains between two sublattice moments
$\mib{M}_1$ and $\mib{M}_2$ shown in Fig.~\ref{fig1} (b) 
even when 3D long-range order is well established in each sublattice.
Note that either $M_{\rm S}$ or $M_{\rm S}^{(xy)}$ in Sec.~3.2 
does not give information
about the relative angle between $\mib{M}_1$ and $\mib{M}_2$.

Here, we measure the collinearity by
$C_{12} = \langle \cos^2 \theta_{12} \rangle$, 
where $\theta_{12}$ describes the angle between $\mib{M}_1$ and $\mib{M}_2$.
Figure~\ref{fig4} shows our MC results of $C_{12}$ for the model (\ref{eq:H}).
$C_{12}$ rapidly increases below $T_{\rm N}$ as shown in the inset. 
System-size dependences of $C_{12}$ 
show that the data approach $1$ in the thermodynamic limit below $T_{\rm N}$,
which indicates that the magnetic order is collinear.

The collinear ordering is understood 
by the so-called order-by-disorder mechanism.
\cite{Villain1977}
Our MC results show that the frustration remaining in the mean-field level 
is reduced by thermal fluctuations and 
the collinear AFM state is stabilized. 
In Ref.~\ref{Tsunetsugu}, the authors have shown that 
quantum fluctuations also favor the collinear state 
to minimize the zero point energy of magnons. 
It is well known that both thermal and quantum fluctuations favor 
a collinear state in many frustrated spin systems. 
This is also the case in the present system. 

The magnetic structure obtained by MC calculations 
is shown in Fig.~\ref{fig1} (b).
The ordering pattern is consistent with 
the experimental result by the neutron scattering.
\cite{Niziol1973}

\section{Summary}

We have found that the interplay between spin and orbital is 
crucial in the geometrically-frustrated vanadium spinels. 
The orbital ordering introduces spatial anisotropy 
in the effective spin exchange interactions and 
drastically reduces the magnetic frustration. 
In the orbital ordered phase, the system can be regarded as 
weakly coupled 1D AFM chains. 
By applying the finite-size scaling analysis to numerical data, 
we have shown that the AFM transition belongs to 
the universality class of the 3D unfrustrated Heisenberg model. 
We have also pointed out the importance of thermal fluctuations 
to stabilize the collinear AFM state by the order-by-disorder mechanism.

\section*{Acknowledgment}

This work was supported by a Grant-in-Aid and NAREGI 
from the Ministry of Education, Science, Sports, and Culture. 
A part of the work was accomplished during Y. M. was staying 
at the Yukawa Institute of Theoretical Physics, Kyoto University, 
with the support from The 21st Century for Center of Excellence program, 
`Center for Diversity and Universality in Physics'.

\end{document}